\documentclass[aps,twocolumn,prl,showpacs,superscriptaddress]{revtex4}
\usepackage{amsmath}
\usepackage{amssymb}
\usepackage{graphicx}
\usepackage{colordvi}
\usepackage{color}

\begin{document}

\title{Competition of resonant and nonresonant paths in resonance-enhanced two-photon single ionization of He by an ultrashort extreme-ultraviolet pulse}
\author{Kenichi L. Ishikawa}
\email[Electronic address: ]{ishiken@atto.t.u-tokyo.ac.jp}
\affiliation{Photon Science Center, Graduate School of Engineering, The University of
Tokyo, 7-3-1 Hongo, Bunkyo-ku, Tokyo 113-8656, Japan}

\author{Kiyoshi Ueda}
\affiliation{Institute of Multidisciplinary Research for Advanced Materials, Tohoku University, Sendai 980-8577, Japan}

\date{\today}

\begin{abstract}
We theoretically study the pulse-width dependence of the photoelectron angular distribution (PAD) from the resonance-enhanced two-photon single ionization of He by femtosecond ($\lesssim 20$ fs) extreme-ultraviolet pulses,
based on the time-dependent perturbation theory and simulations with the full time-dependent Schr\"{o}dinger equation. 
In particular, we focus on the competition between resonant and nonresonant ionization paths, which leads to the relative phase $\delta$ between the $S$ and $D$ wave packets distinct from the corresponding scattering phase shift difference. When the spectrally broadened pulse is resonant with an excited level, the competition varies with pulse width, and, therefore, $\delta$ and the PAD also change with it. On the other hand, when the Rydberg manifold is excited, $\delta$ and the PAD do not much vary with the pulse width, except for the very short-pulse regime.
\end{abstract}

\maketitle

Multiphoton ionization of atoms has consistently been receiving a great deal of attention for decades (see e.g.\cite{Gontier1971PRA,Beers1975PRA,Crance1977PRA,Andrews1977JPB,McClean1978JPB,McClean1979JPB, Jackson1982JPB, Dixit1983PRA, He2002PRA, Selstoe2007PRA,Varma2009PRA}). The advent of intense extreme-ultraviolet (EUV) sources such as high-harmonic generation (HHG) and free-electron lasers (FEL) has enabled two-photon ionization (TPI) of species with a deep ionization potential such as He \cite{Hasegawa2005PRA,Kobayashi1998OL,Mitzner2009PRA,Sorgenfrei2010RSI,Moshammer2011OE} and ${\rm N}_2$ \cite{Nabekawa2006PRL}. 
Upon photoionization, the continuum electron wave packet is emitted, which is a superposition of different partial waves, each with its own orbital angular momentum, intensity, and phase. Photoelectron angular distribution (PAD), nowadays extensively studied by the velocity map imaging technique (see e.g. \cite{Haber2009PRA,Rouzee2011PRA}), contains information on the interference of these different partial waves.

In this Letter, we theoretically study the pulse-width-dependence of the PAD from two-photon single ionization of He by femtosecond (fs) EUV pulses. Especially, we focus on situations where the pulse is closely resonant with an excited level, i.e., resonance-enhanced TPI. We have chosen He as a target atom for the following reasons: first, its single-electron excitation energies, e.g., 21.218 eV for $1s2p\, ^1P$ and 23.087 eV for $1s3p\, ^1P$ \cite{NIST}, coincide with the 13th and 15th harmonic photon energies of a Ti:Sapphire laser, respectively, and also with the typical wavelength range of EUV FELs such as the Spring-8 Compact SASE Source (SCSS) \cite{Shintake2008NPhoton}, the Free-electron LASer at Hamburg (FLASH) \cite{Ackermann2007NPhoton},  and FERMI \cite{FERMIwebsite}. Second, its simple electronic structure allows for exact time-dependent numerical analysis \cite{ATDI2005,Parker2001JPB,Pindzola1998PRA,Pindzola1998JPB,Colgan2001JPB}, in great contrast to alkali atoms. 

In the case of resonance-enhanced TPI, the resonant ionization path via resonant levels and the nonresonant path via nonresonant intermediate levels coexist \cite{Beers1975PRA}. Our results show that in the few fs regime, the competition between the two paths can be controlled by changing the pulse width when the pulse is resonant with a single excited level. The relative phase $\delta$ between the different partial waves ($S$ and $D$ for He) would be just the scattering phase shift difference for nonresonant TPI and resonant $(1+1^\prime)$ TPI \cite{Haber2009PRA}. For the present case, on the other hand, the ionization-path competition give rise to an additional contribution $\delta_{ex}$. This contribution clearly manifests itself in the PAD, and both the PAD and $\delta$ vary with the pulse width. When the pulse becomes so short that its spectrum gets broader than the level spacing and resonant with multiple levels, especially the Rydberg manifold, $\delta$ is still different from the scattering phase shift difference, but does not vary with $T$.
We further explore how the chaotic nature of FEL radiation \cite{Mitzner2008OE,Schlotter2010OL,Bonifacio1994PRL,Saldin1998OC,Krinsky2003} affects the PAD. Our analysis using the partial-coherence method \cite{Pfeifer2010OL} indicates that 
the PAD is between those corresponding to the coherence time and the mean pulse duration.


Let us first do a simple analysis on how the relative importance of the resonant and nonresonant paths depends on pulse width, based on the second-order time-dependent perturbation theory within the common rotating wave approximation. The dynamic Stark effect is negligible for pulse parameters used in the present study. We consider the process where a laser pulse with a central frequency $\omega$ and a pulse envelope $f(t)$, linearly polarized in the $z$ direction, promotes an atomic electron from an initial state $| i \rangle$ to a final continuum state $| f \rangle$ through two-photon absorption. The complex amplitude $c_f$ of the final state after the pulse in the interaction picture can be written as,
\begin{equation}
\label{eq:SOTDPT}
c_f=\sum_m\int_{-\infty}^{\infty}\mu_{fm}e^{i\Delta_f t}f(t)\left(\int_{-\infty}^t \mu_{mi} e^{i\Delta_m t^\prime}f(t^{\prime})dt^\prime\right)dt,
\end{equation}
where $\mu_{mn}$ denotes the dipole transition matrix element between states $m$ and $n$, $\Delta_m = \omega_m-(\omega_i+\omega)$, $\Delta_f = \omega_f-(\omega_m+\omega)$ with $\omega_m$ being the energy eigen-value of state $m$, and the sum runs over all the intermediate bound and continuum states $m$. Although a rectangular pulse is often assumed in previous work \cite{Beers1975PRA}, we take, as a more realistic choice, a Gaussian profile $f(t) = E_0 e^{-t^2/2 T^2}$, with $E_0$ and $T$ being the field amplitude and the pulse width, respectively. For the case of $\omega_f=\omega_i+2\omega$, in particular, one can perform the integrals in Eq.\ (\ref{eq:SOTDPT}) analytically to obtain a physically transparent expression:
\begin{equation}
\label{eq:SOTDPT-Gaussian}
c_f=\pi E_0^2T^2\sum_m\,\mu_{fm}\mu_{mi}\left[e^{-{\Delta_m^2T^2}}-i\frac{2}{\sqrt{\pi}}\,F\left(\Delta_m T\right)\right],
\end{equation}
where $F(x)$ denotes Dawson's integral \cite{NIST-math}, which tends to $x$ near the origin and $1/2x$ for $x\to\infty$. Only resonant states within the spectral width of the pulse contribute to the first term, corresponding to the resonant path. On the other hand, the asymptotic behavior of $F(x)$ suggests that all the intermediate states except for the exact resonance ($\Delta_m=0$) participate in the second term, as expected for nonresonant paths. While either term dominates for a relatively long pulse (ps and ns), we can expect that the two terms are comparative for ultrashort ($\sim$ a few fs) pulses and that their relative importance, which may be expressed as $\arg c_f$, varies with $T$. In such a situation, the amplitude ratio $c_{S}/c_{D}$ between the final $S$ and $D$ continuum states is complex, since the branching ratio $\mu_{Sm}/\mu_{Dm}$ of the transitions from the intermediate $P$ states $m$ to each state depends on $m$. While the actual outgoing wave packets involve the contribution from the final states with $\omega_f\ne\omega_i+2\omega$, it is instructive to write $\arg c_S/c_D$ using Eq.\ (\ref{eq:SOTDPT-Gaussian}) as follows:
\begin{equation}
\label{eq:amplitude_ratio}
\frac{c_S}{c_D}=\frac{\mu_{Sr}}{\mu_{Dr}} \frac{\sqrt{\pi}Te^{-\Delta_r^2T^2}-i[a_S+2F(\Delta_r T)T]}{\sqrt{\pi}Te^{-\Delta_r^2T^2}-i[a_D+2F(\Delta_r T)T]},
\end{equation}
with $a_f (f=S,D)=(\mu_{fr}\mu_{ri})^{-1}\sum_{m(\ne r)}\mu_{fm}\mu_{mi}/\Delta_m$. Here we have assumed that only one intermediate state $r$ is resonant with the pulse and that $F(\Delta_m T)\approx (2 \Delta_m T)^{-1}$ for all the other intermediate states. Hence, the competition between the resonant and nonresonant paths affects the interference between the outgoing $S$ and $D$ wave packets and manifests itself in the photoelectron angular distribution.

The photoelectron angular distribution is given by \cite{Smith1988AAMP},
\begin{equation}
\label{eq:pad}
I(\theta)=\frac{\sigma}{4\pi}\left[1+\beta_2P_2(\cos\theta)+\beta_4P_4(\cos\theta)\right],
\end{equation}
where $\sigma$ is the total cross section, $\theta$ is the angle between the laser polarization and the electron velocity vector, and $\beta_2$ and $\beta_4$ are the anisotropy parameters associated with the second- and fourth-order Legendre polynomials, respectively. The interference of the $S$ and $D$ wave packets produces a photoelectron angular distribution proportional to $\left| |c_S|e^{i\delta_0}Y_{00} - |c_D|e^{i\delta_2}Y_{20}\right|^2$, with $\delta_l$ being the phase of the partial wave, or the {\it apparent} phase shift. Then, the anisotropy parameters can be described by,
\begin{equation}
\label{eq:beta2and4}
\beta_2=\frac{10}{W^2+1}\left[\frac{1}{7}-\frac{W}{\sqrt{5}}\cos\delta\right], \quad
\beta_4=\frac{18}{7(W^2+1)},
\end{equation}
%
%
where $W=|c_S/c_D|$ and $\delta = \delta_0-\delta_2$ \cite{chuuigaki}. The apparent phase shift difference $\delta=\delta_{sc}+\delta_{ex}$ consists of a part $\delta_{sc}$ intrinsic to the continuum eigen wave-functions (scattering phase shift difference), which has previously been studied both theoretically \cite{Oza1986PRA,Chang1995PRA,Gien2002JPB} and experimentally \cite{Haber2009PRA}, and the extra contribution $\delta_{ex}=\arg c_S/c_D$ from the competition of the two paths. This situation present a contrast to the case of the photoionization from photo-excited states \cite{Haber2009PRA}, where the nonresonant path is absent and only $\delta_{sc}$ is present ($\delta=\delta_{sc}$). 

We now verify this qualitative idea, using direct numerical solution of the full-dimensional two-electron time-dependent Schr\"odinger equation (TDSE) \cite{ATDI2005}:
%
\begin{equation}
\label{eq:TDSE_He}
i\frac{\partial\Phi ({\bf r}_1,{\bf r}_2,t)}{\partial t} =[H_0+H_I(t)]\Phi ({\bf r}_1,{\bf r}_2,t),
\end{equation}
with the atomic and interaction Hamiltonian,
\begin{equation}
\label{eq:atomic_hamiltonian}
H_0 = -\frac{1}{2}\nabla_1^2 -\frac{1}{2}\nabla_2^2-\frac{2}{r_1}-\frac{2}{r_2}+\frac{1}{|{\bf r}_1-{\bf r}_2|},
\end{equation}
\begin{equation}
\label{eq:interaction_hamiltonian}
H_I(t) = (z_1+z_2)f(t)\sin\omega t.
\end{equation}
We solve Eq.\ (\ref{eq:TDSE_He}) using the time-dependent close-coupling method \cite{ATDI2005,Parker2001JPB,Pindzola1998PRA,Pindzola1998JPB,Colgan2001JPB}. The numerically obtained excitation energies for the $1s2p\, ^1P$ and $1s3p\, ^1P$ states are 21.220 and 23.086 eV, respectively, in fair agreement with the experimental values (21.218 and 23.087 eV \cite{NIST}, respectively). Sufficiently after the pulse has ended, we calculate $\beta_2$ and $\beta_4$ by integrating the ionized part of $|\Phi ({\bf r}_1,{\bf r}_2)|^2$ over $r_1,r_2,\theta_2,\phi_1,\phi_2$, from which one obtains $W$ and $\delta$ by solving Eq.\ (\ref{eq:beta2and4}). We use the values of $\delta_{sc}$ from \cite{Gien2002JPB} to calculate $\delta_{ex}=\delta-\delta_{sc}$. The calculation has been done for a Gaussian pulse envelope with a peak intensity of $10^{11}\,{\rm W/cm}^2$, at which we have confirmed that the interaction is still in the perturbative regime. 

\begin{figure}[tb] 
   \centering
   \includegraphics[width=83mm]{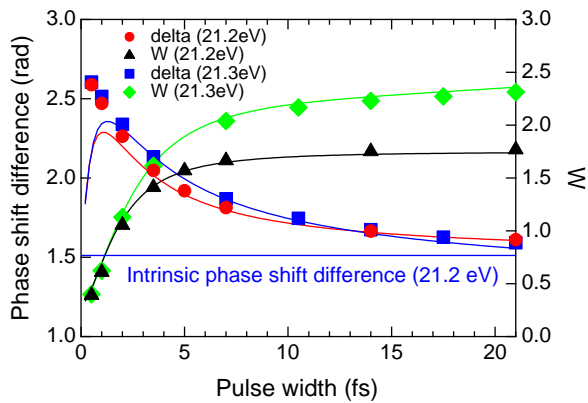} 
   \caption{(Color online) The pulse-width dependence of the TDSE-derived apparent phase shift difference (relative phase) $\delta$ (left axis) and $W=|c_s/c_d|$ (right axis) for $\hbar\omega=21.2\,{\rm eV}$ and $21.3\,{\rm eV}$. The thin horizontal line denotes the value (1.511 \cite{Gien2002JPB}) of the intrinsic scattering phase-shift difference $\delta_{sc}$ for $\hbar\omega=21.2\,{\rm eV}$. That for $\hbar\omega=21.3\,{\rm eV}$ is 1.491 \cite{Gien2002JPB}. Solid lines are the results of fitting using Eq.\ (\ref{eq:amplitude_ratio}).}
   \label{fig:01}
\end{figure}

\begin{figure}[tb] 
   \centering
   \includegraphics[width=83mm]{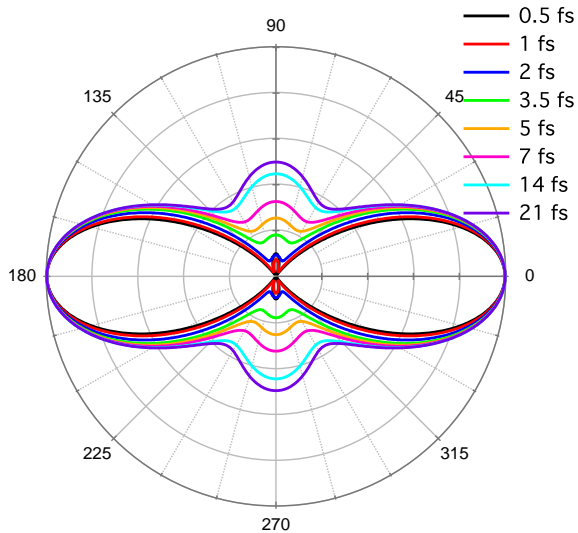} 
   \caption{(Color online) The pulse-width dependence of the photoelectron angular distribution for $\hbar\omega=21.2\,{\rm eV}$.}
   \label{fig:02}
\end{figure}

The pulse-width dependence of $\delta$ and $W$ for $\hbar\omega=21.2\,{\rm eV}$ and $\hbar\omega=21.3\,{\rm eV}$ close to the $1s2p$ resonance (21.218 eV) is shown in Fig.\ \ref{fig:01}. The calculations have been done at different values of full-width-at-half-maximum (FWHM) pulse width $T_{1/2}=2\sqrt{\ln 2}\,T$ between 500 as and 21 fs. As expected, both $\delta$ and $W$ substantially changes with pulse width, especially when the pulse is shorter than 10 fs. Accordingly, the PAD also varies as shown in Fig.\ \ref{fig:02}. One finds that the distribution to the direction perpendicular to the laser polarization, i.e., $\theta\approx 90^\circ, 270^\circ$ decreases as the pulse is shortened. This can be understood as follows: roughly speaking, $\delta$ changes from $\sim \frac{\pi}{2}$ to $\sim \pi$ as $T_{1/2}$ varies from 21 fs to 500 as. Thus, $c_s/c_d$ is approximately real and negative in the short-pulse limit, which leads to the cancellation between $Y_{00}(\theta,\varphi)$ and $Y_{20}(\theta,\varphi)$ around $\theta=\frac{\pi}{2}$. 
As stated earlier, strictly speaking, Eq.\ (\ref{eq:amplitude_ratio}) is applicable only to $\omega_f=\omega_i+2\omega$, and the actual PAD involves integration over $\omega_f$. Nevertheless, the results in Fig.\ \ref{fig:01} can well be described by Eq.\ (\ref{eq:amplitude_ratio}) (solid lines in Fig.\ \ref{fig:01}), except for $\delta$ in the ultrashort pulse regime $T_{1/2}\lesssim 1\,{\rm fs}$, where the spectrum becomes broader than the level spacing.

\begin{figure}[tb] 
   \centering
   \includegraphics[width=83mm]{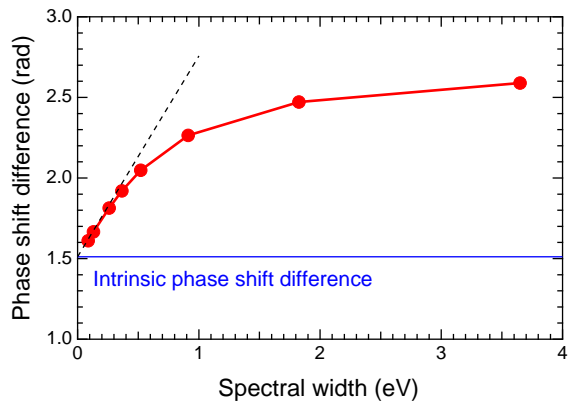} 
   \caption{(Color online) The spectral-width dependence of the TDSE-derived apparent phase shift difference (relative phase) $\delta$ for $\hbar\omega=21.2\,{\rm eV}$. The thin horizontal line denotes the value (1.511 \cite{Gien2002JPB}) of $\delta_{sc}$. Thin dashed line plots the asymptotic behavior Eq.\ (\ref{eq:linear}).}
   \label{fig:03}
\end{figure}

With increasing pulse duration, $\delta$ approaches the scattering phase shift difference $\delta_{sc}$, and the PAD changes only slowly with $T_{1/2}$ (Fig.\ \ref{fig:02}). When the pulse is resonant ($\Delta_r T \ll 1$) and sufficiently long ($T\gg a_S, a_D$) at the same time, assuming that the resonant excitation is not saturated, one can approximate the extra phase shift as 
\begin{equation}
\label{eq:linear}
\delta_{ex}\approx(a_D-a_S)/\sqrt{\pi}\,T, 
\end{equation}
hence, it is proportional to the spectral width, which can be confirmed in Fig.\ \ref{fig:03}.

\begin{figure}[tb] 
   \centering
   \includegraphics[width=83mm]{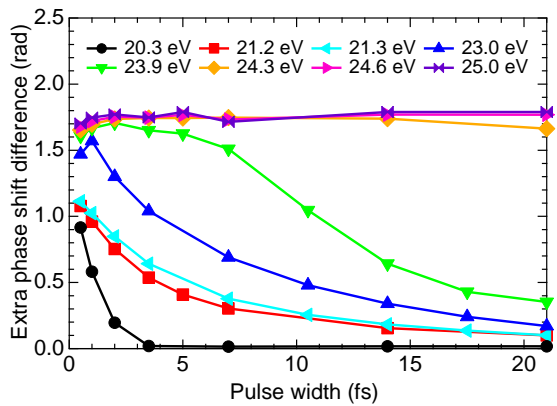} 
   \caption{(Color online) Pulse-width dependence of $\delta_{ex}$ for different values of photon energy. The resonant excitation energies for $1s2p^1P$, $1s3p^1P$, $1s4p^1P$, $1s5p^1P$ levels are 21.218, 23.087, 23.742, and 24.046 eV \cite{NIST}.}
   \label{fig:04}
\end{figure}

On the other hand, if we plot $\delta$ as a function of spectral width (Fig.\ \ref{fig:03}), $\delta$ tends to an asymptotic value in the wide-spectrum, i.e., short-pulse limit. Correspondingly, the PAD does not change much with the pulse width for $T_{1/2}\lesssim 1\,{\rm fs}$ (Fig.\ \ref{fig:02}). This is because the pulse becomes resonant with multiple levels; the spacing between the $1s2p$ and $1s3p$ is 1.9 eV. Indeed, one can show that when many neighboring states are resonantly excited by the pulse, the extra phase shift difference $\delta_{ex}$ does not much depend on the pulse duration. This especially applies when the photon energy lies in the Rydberg manifold, and exceeds the ionization potential (24.59 eV), i.e., in the case of above-threshold two-photon ionization.
In Fig.\ \ref{fig:04} we compare the pulse-width dependence of $\delta_{ex}$ for different values of $\hbar\omega$. While $\delta\approx\delta_{sc}$ ($\delta_{ex}\approx 0$) for nonresonant pulses ($\hbar\omega=20.3\,{\rm eV}$, $T_{1/2}\gtrsim 3.5\,{\rm fs}$), when the pulse is close to resonance with an excited level ($\hbar\omega=21.2, 21.3$, and $23.0\,{\rm eV}$), $\delta_{ex}$ rapidly changes with $T_{1/2}$. On the contrary, $\delta_{ex}$ is nearly constant for $\hbar\omega=24.3, 24.6$, and $25.0\,{\rm eV}$. At $T_{1/2}\lesssim 2$ fs, the spectrum is so broad that $\delta_{ex}$ restarts to change slightly. One also sees that the transition across the ionization potential is smooth. It should be pointed out that the extra phase shift difference due to free-free transitions plays a significant role in the recently observed time delay in photoemission by attosecond EUV pulses \cite{Schultze2010Science,Kluender2011PRL}. 

\begin{figure}[tb] 
   \centering
   \includegraphics[width=83mm]{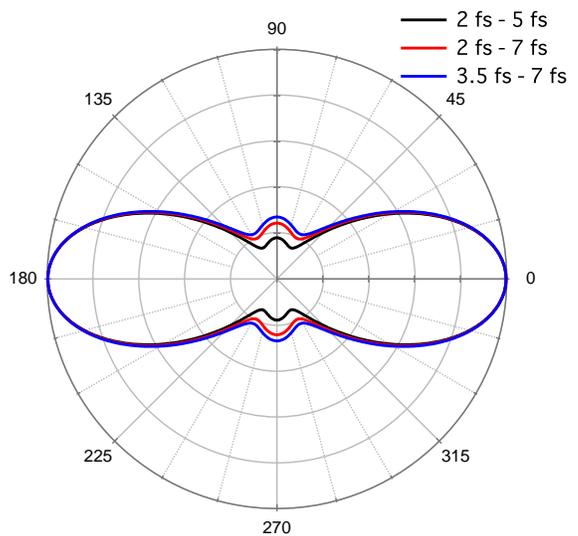} 
   \caption{(Color online) Photoelectron angular distribution by chaotic pulses for $\hbar\omega=21.2\,{\rm eV}$ and three pairs of (CT)-(MPW) indicated in the legend.}
   \label{fig:05}
\end{figure}

Coherent EUV pulses considered so far may be generated by HHG sources and HHG-seeded FELs \cite{Togashi2011OE}. As is well known, on the other hand, the temporal pulse shapes of FEL operating in the self-amplified spontaneous emission (SASE) mode fluctuate from shot to shot \cite{Mitzner2008OE,Schlotter2010OL,Bonifacio1994PRL,Saldin1998OC,Krinsky2003}. In order to investigate the effects of the chaotic nature, we have performed numerical experiments for EUV pulses randomly generated by the partial-coherence method \cite{Pfeifer2010OL}. The obtained values of $W$ and $\delta$ for several combinations of  coherence time (CT) and mean pulse width (MPW), and the corresponding PAD are shown in Table \ref{table:pcp} and Fig.\ \ref{fig:05}, respectively. We can see from the table that $\delta$ and $W$ take values between those corresponding to the CT and the MPW in most cases. Also, the PAD is of a shape between those for the CT and the MPW. The detailed mechanism underlying these somewhat empirical findings will require further investigation taking photon statistics into account.

\begin{table}
  \centering 
  \caption{$W$ and $\delta$ calculated for chaotic pulses generated by the partial-coherence method for several pairs of coherence time (CT) and mean pulse width (MPW). The average values and standard deviation errors of $W$ and $\delta$ by 48 runs are listed. The rows with the same CT and MPW values are for fully coherent pulses as in Figs.\ \ref{fig:01} and \ref{fig:04}.}\label{table:pcp}
  \begin{tabular}{ccccc}
\hline
\hline\hline
$\hbar\omega$ (eV) & CT (fs) & MPW (fs) & $W$ & $\delta$ \\
\hline
21.2 & 2    & 5 & $1.31\pm 0.08$ & $2.03\pm 0.04$ \\
         & 2    & 7 & $1.42\pm 0.09$ & $1.91\pm 0.04$ \\
         & 3.5 & 7 & $1.49\pm 0.09$ & $1.87\pm 0.03$ \\
         & 2    & 2 & 1.05 & 2.26 \\
         & 3.5 & 3.5 & 1.41 & 2.05 \\
         & 5    & 5 & 1.57 & 1.92 \\
         & 7    & 7 & 1.67 & 1.81 \\
23.0 & 2 & 7 & $0.645\pm 0.061$ & $2.14\pm 0.08$ \\
         & 3.5 & 7 & $0.890\pm 0.058$ & $2.15 \pm 0.06$ \\
         & 2 & 2 & 0.464	 & 2.71 \\
         & 3.5& 3.5 & 0.679 & 2.45 \\
         & 7 & 7 & 1.08 & 2.10 \\         
\hline\hline
\end{tabular}
\end{table}

In summary, due to the competition between the resonant and nonresonant ionization paths, the relative phase $\delta$ between the $S$ and $D$ photoelectron wave packets from the resonance-enhanced two-photon ionization of He by fs EUV pulses is different from the scattering phase shift difference $\delta_{sc}$ which would be expected for single-photon ionization \cite{Haber2009PRA} and nonresonant two-photon ionization, and rapidly changes with the pulse width when the pulse is resonant with an intermediate excited state and $2\,{\rm fs} \lesssim T_{1/2} \lesssim 10\,{\rm fs}$. Accordingly, the photoelectron angular distribution varies with $T_{1/2}$ as well. 
Also, $\delta_{ex}$ is finite but constant independent of $T_{1/2}$ when the Rydberg manifold is excited
 . Hence, the control of the competition between the resonant and nonresonant paths in He by pulse width is a unique feature of a-few-fs EUV pulses. The PAD is affected by the chaotic nature of SASE FEL pulses, and takes a shape between those corresponding to the coherence time and the mean pulse width of the pulses. The results of the present study stress the importance of the account of the nonresonant paths in the interpretation of resonant two-photon, single- or two-color, ionization experiments by state-of-the-art ultrashort EUV sources.    

\begin{acknowledgments}
We wish to thank N. Kabachnik for fruitful discussions. 
K.L.I. gratefully acknowledges support by the APSA Project (Japan), KAKENHI (No. 23656043 and No. 23104708), the Cooperative Research
Program of ``Network Joint Research Center for Materials and Devices,'' (Japan) and the Project of Knowledge Innovation Program (PKIP) of Chinese Academy of Sciences (Project No. KJCX2.YW.W10). K.U.acknowledges support by ``promotion of X-ray Free Electron Laser Research'' from MEXT, the Management Expenses Grants for National Universities Corporations from MEXT, and IMRAM research program.
\end{acknowledgments}


\end{document}